\documentclass[aps,prl,showpacs,floatfix,twocolumn,superscriptaddress]{revtex4}
\usepackage{amssymb} \usepackage{amsmath}
\usepackage{graphicx}
\usepackage[latin1]{inputenc}
\def\be{\begin{equation}}
\def\ee{\end{equation}}
\def\bea{\begin{eqnarray}}
\def\eea{\end{eqnarray}}
\def\ba{\begin{array}}
\def\ea{\end{array}}

\begin{document}
\setlength{\parindent}{0pt} \newcommand{\llangle}{\langle \langle}
\newcommand{\rrangle}{\rangle \rangle}
\title{Temperature dependence of the conductivity of ballistic graphene}

\author{Markus M{\"u}ller}
\affiliation{The Abdus Salam International Center for Theoretical Physics, Strada Costiera 11, 34014 Trieste, Italy}
\author{Matthias Br{\"a}uninger}
\affiliation{Institute for Theoretical Physics and Astrophysics, University of
W{\"u}rzburg, D-97074 W{\"u}rzburg, Germany}
\author{Bj{\"o}rn Trauzettel}
\affiliation{Institute for Theoretical Physics and Astrophysics, University of
W{\"u}rzburg, D-97074 W{\"u}rzburg, Germany}


\begin{abstract}
We investigate the temperature dependence of the conductivity in ballistic graphene using Landauer transport theory.
We obtain results which are qualitatively in agreement with many features recently observed in transport measurements on high mobility suspended graphene. The conductivity $\sigma$ at high temperature $T$ and low density $n$ grows linearly with $T$, while at high $n$ we find $\sigma\sim \sqrt{|n|}$ with negative corrections at small $T$ due to the $T$-dependence of the chemical potential. At moderate densities the conductivity is a non-monotonic function of $T$ and $n$, exhibiting a minimum at $T=0.693 \hbar v\sqrt{|n|}$ where $v$ is the Fermi velocity. We discuss two kinds of Fabry-Perot oscillations in short nanoribbons and their stability at finite temperatures.
\end{abstract}

\date{\today} \maketitle
Ballistic transport in graphene has been theoretically analyzed \cite{Kat06,Two06,Sch08} shortly after the pioneering measurements of the quantum Hall effect in the first single-layer graphene samples \cite{Nov05,Zha05}. So far theoretical approaches concentrated on low temperatures and density,
predicting a finite and universal minimum conductivity $\sigma=(4/\pi) e^2/h$ as well as a universal Fano factor $F\equiv S/2eI=1/3$ (the ratio of the current noise $S$ and the average current $I$, $e>0$ being the electron charge). This is a surprising result in the ballistic transport regime where  one might naively expect no current noise at all. The reason for finite noise lies in the dynamics of charge carriers in graphene which at low energies is governed by the massless Dirac equation and not by the massive Schr{\"o}dinger equation.
As a result, the ballistic transport 
resembles diffusive transport in a normal metal, which has been coined {\it pseudo-diffusive transport} in the literature.
Both the predictions for the conductivity as well as the Fano factor have been experimentally observed \cite{Mia07,Dic08,Dan08}.

Recently, mobilities approaching 200000 cm$^2$/Vs have been reported for ultraclean suspended graphene \cite{Bol08,Du08} which are an order of magnitude larger than typical mobilities of graphene deposited on a substrate. The transport characteristics of these experiments suggest that the samples reach the ballistic regime with respect to disorder scattering. Indeed, the conductance scales with the number of channels, which is inconsistent with dominant scattering from charged impurities, or ripples~\cite{Ripples}, the latter being the most likely elastic scatterers in suspended graphene.
Further, $\sigma$ is proportional to the sample length $L$ for small sizes and low temperature $T$.

While these two features suggest ballistic transport, they do not actually rule out the presence of inelastic scattering due to electron-electron interactions, which preserve both the density dependence $\sigma\sim \sqrt{n}$ and the proportionality to $L$~\cite{MuellerGlazman} - except for $n=0$ and finite $T$ where interactions lead to a $L$-independent finite conductivity, in contrast to the ballistic case (see Eq.~(\ref{sigmaTmu=0}) below). The data of Ref.~\cite{Bol08} is indeed likely to bear fingerprints of Coulomb interactions in a certain parameter regime close to the neutrality point and at higher temperature where interactions are strongest~\cite{Mue08}. On the other hand, however, the significant linear increase of $\sigma(T)$ at charge neutrality reported in Ref.~\cite{Du08} is inconsistent with dominant electron-electron interactions. (In the experiment~\cite{Du08} the latter are presumably screened by nearby metallic electrodes.)
This insulating trend in the temperature dependence, and the opposite metallic trend at moderate density, which is reported in both Refs.~\cite{Bol08,Du08}, has remained a puzzle.
The authors of Ref.~\cite{Bol08} attempted to explain the latter by scattering from phonons. However, they point out that the decrease of the effect with density is not consistent with such a scenario. Ascribing the observed effects to electron-electron interactions is not consistent either, since those would exhibit a rather weak temperature dependence~\cite{Fri08,Mue08}, with a metallic instead of an insulating trend at low density.
Hwang and Das Sarma~\cite{Hwa08} have proposed scattering from charged impurities and their temperature dependent screening as a possible explanation for the different behavior at low and high density. However, this model cannot account for the density dependence of the data and the length independence of the conductance at low temperature.


In this Letter, we analyze impurity-free transport at finite $T$. As motivated above, we assume interactions to be weak, either due to dielectric attenuation from a substrate, or screening by nearby metals or the finite density of electrons themselves. We show that the extension of the ballistic transport model of Ref.~\cite{Two06} to $T>0$ qualitatively explains most of the features observed in Ref.~\cite{Du08} and, apart from the low $n$ and high $T$ regime, also those in Ref.~\cite{Bol08}. In particular, in the absence of interactions, the minimum conductivity is shown to grow linearly with $T$.
At finite carrier density $n$, the initial $T$ dependence is negative, but changes sign at a temperature of order $T\sim \hbar v \sqrt{|n|}$ where $v \approx 10^6$ m/s is the Fermi velocity.

In the model for ballistic graphene proposed in Ref.~\cite{Two06}, a nanoribbon of width $W$ is suspended between left and right reservoirs (wide graphene regions), which are a distance $L$ apart. The leads and the sample region are subject to a step-like electrostatic potential 
\begin{equation} \label{mu}
\phi(x) = \left\{ \begin{array}{ccc} \phi_\infty & x < 0 & (\text{left lead}), \\ 0 & 0 < x < L & (\text{sample}), \\ \phi_\infty & x > L & (\text{right lead}), \end{array} \right.
\end{equation}
the zero of energy being chosen as the Dirac point in the sample region. Modeling the leads by highly doped graphene ($|\phi_\infty|$ much larger than the gate induced Fermi level in the sample) one finds that the parameter $\phi_\infty$ drops out in the end.

Ballistic transport requires the sample length to be shorter than the mean free path $\ell$. In nearly impurity free samples 
the latter is limited by inelastic scattering, dominated at low $T$ by Coulomb scattering, $\ell\approx (\hbar v/\alpha^{2})T^2/{\rm max}(|\mu|, T)$ ~\cite{Fri08,Mue08}. The latter is strongest in the non-degenerate regime $T>|\mu|$, but is relatively weak outside because of screening effects, and the down-renormalization of the Coulomb coupling constant $\alpha$. The latter is of order $O(1)$ in the unscreened situation pertaining to free-hanging graphene in the non-degenerate regime. However, on a clean substrate with large dielectric constant, or in the presence of nearby metallic contacts, as in Ref.~\cite{Du08}, the effective value of $\alpha$ is substantially reduced. This opens a reasonably large window of applicability for the non-interacting theory which we develop below.

The zero temperature conductivity at a fixed Fermi level $E$ in the sample is given by the Landauer formula
\begin{equation}
\label{sigma}
\sigma_0(E) = \frac{L}{W} G(E) = \frac{L}{W} \frac{g e^2}{h} \sum_{n=0}^{\infty} T_n (E),
\end{equation}
where $G(E)$ is the conductance, $g=4$ is the degeneracy due to spin and valley degrees of freedom, and $n$ labels the transverse modes of the graphene ribbon. Their transmission probability follows from solving the propagation through the potential (\ref{mu}), 
\begin{equation} \label{tn}
T_n(E) = \frac{E^2 - (\hbar v q_n)^2}{E^2 - (\hbar v q_n)^2 \cos^2({k}_n L)},
\end{equation}
where ${k}_n \equiv (\hbar v)^{-1} \sqrt{E^2-(\hbar v q_n)^2}$.
The transverse momentum $q_n$ is defined for various boundary conditions as $q_n = (n+\gamma) \pi/W$. Below we use  $\gamma=1/2$ corresponding to infinite mass confinement, see Ref.~\cite{Two06}.

The energies $E_{W,L} \equiv \hbar v/\{W,L\}$ set typical scales below which finite size effects are important. At higher energies confinement effects are unimportant. This is easily seen in the asymptotics of the $T=0$ conductivity in the limit $W\gg L$ (which is independent of $\gamma$~\cite{Two06})
%
\bea
\label{sigmaapprox}
&&\sigma\left(y\equiv \frac{|E|}{E_L}\right) =  \frac{4e^2}{\pi h} y\int_0^\infty \frac{(1-u^2)du}{1-u^2\cos^2[y\sqrt{1-u^2}]}\\
&&\quad\quad = \frac{e^2}{h}\cdot \left\{ \begin{array}{l}
\frac{4}{\pi} \,\left[1+0.10094 y^2+O(y^4)\right], \quad y\ll 1\\
y+\frac{\sin(2y-\pi/4)}{2\sqrt{\pi y}}+O(1/y),\quad y\gg 1.
\end{array}
\right.\nonumber
\eea
%
This reproduces the well-known minimal conductivity $\sigma_{\rm min}(\mu=T=0) = \frac{4 e^2}{h} \frac{1}{\pi}$. For $E\gg E_L$, we find $\sigma$ to be essentially proportional to $L$ and $E$, which reflects the number of conducting channels at energy $E$. The oscillatory interference term will be discussed in detail further below.

We now take into account the effects of finite temperature in this model.
In linear response, the Landauer formalism yields the exact formula
for the conductivity
\begin{equation} \label{sigmaT-exact}
\sigma(\mu, T) = T^{-1}\int \sigma_0(E) f(E) [1-f(E)]\, dE,
\end{equation}
where $f(E)= (1 + \exp\left[(E - \mu)/T \right])^{-1}$ is the Fermi distribution function ($k_B \equiv 1$ throughout) and $\mu$ is the chemical potential in the sample.

For $T\gg E_L$ it is justified to neglect the oscillatory term in (\ref{sigmaapprox}), and we find the $T$-dependent conductivity
\bea \label{sigmaT}
\sigma(\mu, T) &\approx& 
\frac{e^2}{h}\frac{L}{\hbar v}\left[|\mu| + 2T\log(1+e^{-|\mu|/T})\right].
\eea

At charge neutrality, and in the non-degenerate case, $T\gg |\mu|$, the conductivity grows linearly with $T$, similarly as observed in Ref.~\cite{Du08},
\begin{equation} \label{sigmaTmu=0}
\sigma(\mu=0, T) \approx  \frac{e^2}{h}\frac{LT}{\hbar v} 2\log(2),
\end{equation}
where we have dropped a small positive offset.
The finite $T$ correction to $\sigma_{\rm min}(T=0)$ is always positive, the result (\ref{sigmaTmu=0}) reflecting the linearly increasing density of states which is sampled at higher temperatures. It is important to note, however,  that this effect does {\em not} survive in the presence of strong electron-electron interactions, for which one would obtain a length independent conductivity at $\mu=0$, with a very small $T$ dependence exhibiting the opposite trend.~\cite{Fri08}



\begin{figure}
  \includegraphics[width=3in]{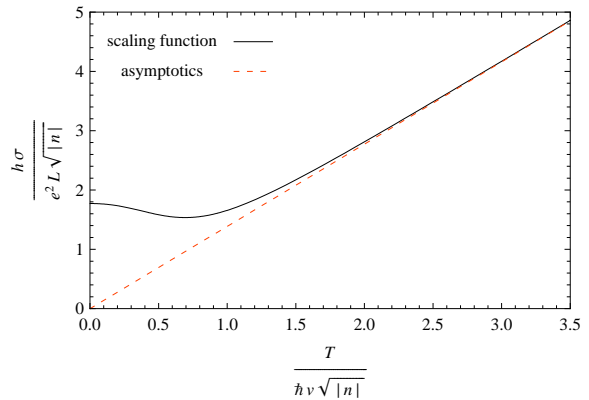}
  \caption{The conductivity as a scaling function of $T/\hbar v |n|^{1/2}$, and its asymptotics at large argument, $\psi(y\gg 1)=2 \log(2) y$. At fixed $n$, $\sigma$ is non-monotonic in temperature, going through a minimum at $T_{\rm min}=0.693 \hbar v|n|^{1/2}$.}
\label{fig:scalingplot}
\end{figure}

So far, we have taken the chemical potential in the sample to be fixed. However, in experiments it is the charge density $n$ which is controlled by the gate potential, rather than the chemical potential \cite{fn1}.
As in the standard Fermi gas, the chemical potential is reduced upon raising the temperature, $\mu(T, n)=\mu_0-(\pi^2/6)T^2/\mu_0$ to lowest order in $T\ll \mu$ in an infinite system. The density $n$ and $\mu_0$ are related by
$\pi(\hbar v)^2n=\mu_0^2$.
In general, in the thermodynamic limit ($W,L\gg \mu_0/\hbar v$) the chemical potential  $\mu(T,n) \equiv T \tilde{\mu}$ satisfies
\bea
\label{muofT}
\int_0^\infty dx\, x \Bigl[ f(x,\tilde\mu) - f(x,-\tilde\mu) \Bigr] = \frac{\pi}{2}\frac{(\hbar v)^2n}{T^2},
\eea
with $f(x,\tilde\mu) = (1+\exp[x-\tilde\mu])^{-1}$. This defines the scaling function
$\tilde{\mu}=\zeta\left(T/\hbar v {|n|^{1/2}}\right)$.
The experimental quantity of interest is the conductivity as a function of $T$ and $n$. From Eqs.~(\ref{sigmaT},\ref{muofT}) one then easily finds the result %
%
%
\bea \label{sigmaT3}
\frac{h}{e^2}\frac{\sigma(n, T)}{L \sqrt{|n|}}
&=&
\psi_\sigma\left(y\equiv \frac{T}{\hbar v \sqrt{|n|}}\right),\\
%
\psi_\sigma(y)&= & y\left[\zeta(y) + 2\log\left(1+e^{-\zeta(y)}\right)\right].
\label{psi}
\eea
The scaling function (\ref{sigmaT3}) is plotted in Fig.~\ref{fig:scalingplot}. It has the asymptotics $\psi(0)=\sqrt{\pi}$ and $\psi(y\gg 1)=2 \log(2) y$, Interestingly, it attains a minimum at $y_{\rm min}=0.6932$ with the value $\psi(y_{\rm min})=1.5356$.
At large density $(\hbar v)^2 |n|\gg T^2$, the conductivity tends to the limit
\bea
\sigma\left(|n|\gg [T/\hbar v]^2\right)= \frac{e^2}{h} L\sqrt{\pi |n|}\,.
\eea
%
At fixed density, the conductivity first decreases as $T$ increases from zero (by a total of  $\sqrt{\pi}-\psi(y_{\rm min})=13.4 \%$)
due to the decrease of the chemical potential. Upon increasing $T$ further, $\sigma$
reaches a minimum at $T_{\rm min}=y_{\rm min}\, \hbar v|n|^{1/2}$ and eventually grows linearly with temperature, approaching the limiting behavior (\ref{sigmaTmu=0}).
The non-monotonicity in the $T$ dependence should be observable for relatively low densities, $n\sim 10^{10} {\rm cm}^{-2}$, for which $T_{\rm min}$ lies within the experimental temperature window, while being low enough for the
ballistic, non-interacting approximation to be applicable, as discussed above.
Such a non-monotonicity has been reported in Fig. 3c of Ref.~\cite{Du08}, whereas it was probably masked by inelastic scattering in Ref.~\cite{Bol08}. At higher densities we predict a slight decrease of conductivity with increasing $T$. We emphasize that the decrease of conductivity results merely from the decrease of $\mu$ without invoking scattering, while the increase of $\sigma$ at higher temperatures reflects the thermal sampling of the higher density of states.
This is to be contrasted with the scenario of Ref.~\cite{Hwa08} where a non-monotonicity was found as well, resulting however from a competition between $T$-dependent screening and thermal sampling of energy dependent scattering rates.



\begin{figure}

\vspace{0.3cm}

\centering
\leavevmode \includegraphics[width=8.0cm]{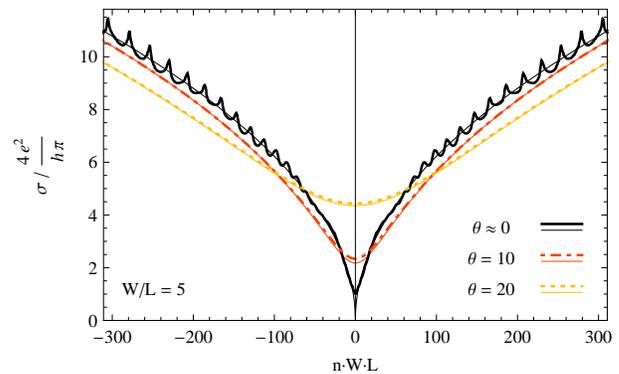}
\caption{(color online) The conductivity $\sigma$ plotted as a function of density $n$ for various temperatures $\theta = T/E_W$. The analytical approximation, neglecting oscillatory terms (thin full lines) is excellent except at low $T$ and $n$. In the latter regime, finite size effects become visible in the form of Fabry-Perot resonances.
[To be compared to Fig. 3c of Ref.~\cite{Du08}.]
}
\label{fig_sig_of_n}
\end{figure}

In order to facilitate the comparison with experimental data, we show in Fig.~\ref{fig_sig_of_n} the dependence of $\sigma$ on the carrier density $n$ for various temperatures $T$, based on the full numerical evaluation of Eq.~(\ref{sigmaT-exact}) for finite $W$. Apart from confirming the above discussed trends, we note a good qualitative agreement with the observations in Ref.~\cite{Bol08}, at least in the regime of larger $n$ where interactions are expected to be weak: the conductivity decreases with $T$, the relative decrease becoming smaller as $n$ increases.

Fig.~\ref{fig_sig_of_n} also illustrates 
the appearance of Fabry-Perot oscillations (FPO) at low temperatures and finite densities, as predicted in Ref.~\cite{Two06}. This effect was not included so far, since we dropped the oscillatory term in (\ref{sigmaapprox}) and approximated the ribbon as infinitely wide.


\begin{figure}

\vspace{0.3cm}

\centering
\leavevmode
\includegraphics[width=8.0cm]{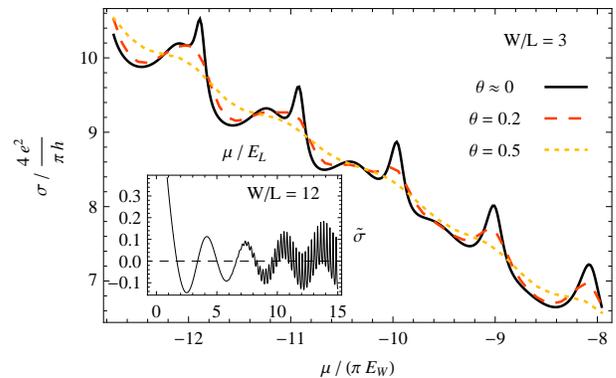}
\caption{(color online) Conductivity 
as a function of chemical potential $\mu$ for various temperatures $\theta\equiv \frac{T}{E_W}$.
The amplitude of the fast Fabry-Perot resonances decays at rather small temperatures $\theta \gtrsim \frac{\delta\mu^{(1)}}{E_W}= \sqrt{2}\frac{(\pi E_W)^2}{\mu^2} \left(\frac{W}{L}\right)^3$. The inset shows the presence of fast and slow oscillations in the deviation from the linear background, $\tilde{\sigma}\equiv \frac{\sigma}{4 e^2/\pi h}-\frac{\pi}{4} \frac{|\mu|}{E_L}$, for a wide sample. The fast oscillations become more pronounced with growing $\mu$, while the slow ones decrease in amplitude.
}
\label{fig_fp}
\end{figure}

As shown in Fig.~\ref{fig_fp}, the FPO's depend on the aspect ratio. Moreover, the peaks are rather easily washed out at finite temperatures.
There are in fact two kinds of FPO's, as one can see from the inset of Fig.~\ref{fig_fp}. Upon tuning the chemical potential or gate voltage, on one hand there are slow oscillations of periodicity $\Delta\mu_s=\pi E_L$, described by the subleading term in (\ref{sigmaapprox}). They originate from the modes which traverse the sample in relatively straight paths, having small transverse quantum numbers $n$. These oscillations are washed out by thermal smearing when $T>\Delta\mu_s$. However, their amplitude is never bigger than $\sqrt{2/11}/\pi\approx 0.14$. Indeed, they can hardly be discerned in Fig.~\ref{fig_sig_of_n}. A second type of oscillations is due to modes which enter the sample nearly at grazing incidence (with largest quantum numbers $n$). Those scatter back and forth between the edges many times when $W\gg L$ (and are thus very sensitive to edge roughness).
Their contribution can be obtained analyzing the transmission factors $T_n$ as a function of chemical potential $\mu$.
Fabry-Perot resonances are expected when
there is a propagating mode at the Fermi level whose longitudinal wavevector $k$ is commensurate with the length of the sample, i.e., for
\bea
k=k_{m} = \frac{m\pi}{L}, \quad m=1,2,\dots\,,
\eea
($m=0$ does not lead to a resonance).
With transverse wavevectors $q_n$, such resonances occur at
\begin{eqnarray}
\label{munm}
\mu_{n}^{(m)} &=& \hbar v \sqrt{q_n^2+k_m^{2}} = \pi E_W  \sqrt{(n+\gamma)^2 +\frac{m^2 W^2}{L^2}}.
\end{eqnarray}
%
One reads off from (\ref{munm}) that the fast oscillations corresponding to fixed $m$ appear with roughly periodic spacings $\Delta\mu_f=\pi E_W$.
In the main panel of Fig.~\ref{fig_fp}, the sharp and soft peaks correspond to $m=1,2$, respectively.

To estimate the width of those peaks at $T=0$ we start from a resonance, $\mu=\mu_n^{(m)}$ and fix $q=q_n$. 
As $\mu$ increases by $\delta\mu$, $k$ changes by $\delta k \approx L \mu \,\delta \mu /(\hbar v)^2 m \pi$.
The peak due to the resonant transmission coefficient $T_n$ [Eq.~(\ref{tn})]
is reduced to half its maximum when $\delta k$ satisfies
\bea
&& 2\left[\mu^2 - (\hbar v q_n)^2\right]=\mu^2 - (\hbar v q_n)^2 \cos^2\left(L(k_{m}+\delta k)\right)\notag\\
&& \quad\quad\quad \approx \mu^2 - (\hbar v q_n)^2 + (\hbar v q_n)^2 (L \,\delta{k})^2/2.
\eea
This translates into a change in chemical potential of
\bea
\label{width}
\delta\mu^{(m)} \approx 
\sqrt{2}(m\pi)^2\frac{E_L^3}{\mu^2} 
\eea
The sharpest peaks correspond to $m=1$. They are well separated form each other if $\delta\mu^{(1)}\ll \Delta\mu_f$, or $\mu/E_L\gg 2^{1/4}(\pi W/L)^{1/2}$, and reach an amplitude close to $1$ at large $\mu$.
These most visible peaks start to broaden and decrease in amplitude when $T>\delta\mu^{(1)}$,  cf., Fig.~\ref{fig_fp}.

Since $\delta\mu^{(1)}$ is a rather low energy scale, and because of the sensitivity of these fast FPO's to edge roughness and inelastic scattering, their experimental observation might prove very challenging.


In summary, the temperature-dependence of the conductivity of ballistic graphene exhibits a rather unexpected behavior. The minimum conductivity always increases with temperature saturating quickly to a linear dependence on $T$ which is a hallmark of weakly interacting systems. At finite density, we predict the conductivity to slightly decrease at low temperatures, but to increase again for $T>T_{\rm min}\sim |n|^{1/2}$.  Finally, we demonstrate two different types of Fabry-Perot oscillations, showing that these fingerprints of ballistic transport are very fragile with respect to finite temperatures.

We thank K.I. Bolotin for interesting discussions. M.M. acknowledges support by the Swiss National Fund for Scientific Research under grants PA002-113151
and PP002-118932, and the Aspen Center of Theoretical Physics for hospitality during the initial stages of this work.  M.B. and B.T. acknowledge financial support by the German Research Foundation (DFG) via grant no. Tr950/1-1.


\begin{thebibliography}{99}
\bibitem{fn1} This is correct if the geometric capacitance per area, $C_{\rm geom} \propto 1/d$ ($d$ being the distance to the gate) is much smaller than the electronic capacitance per area, $C_{\rm el} = e^2d\mu/dn = (2/\pi^{1/2}) (e^2/\hbar v) n^{1/2}$ (at $T=0$).
\end{thebibliography}
\end{document}